\documentclass{aa}
\usepackage{graphicx,psfrag,amssymb,amsmath}
\usepackage{txfonts}
\usepackage{natbib}
\usepackage{epsfig}
\usepackage{caption}
\usepackage{threeparttable}
\bibpunct{(}{)}{;}{a}{}{,}
\begin{document}
\title{Modeling the water line emission from the high-mass star-forming region AFGL\,2591}
\author{D. R.~Poelman 
           \inst{1,2}
           \and
        F. F. S. van der Tak
           \inst{2,1}
                   }
\offprints{D.R.Poelman@astro.rug.nl}
\institute{Kapteyn Astronomical Institute, P.O. Box 800, 9700 AV Groningen, the Netherlands\\
\email{D.R.Poelman@astro.rug.nl}
        \and
         SRON Netherlands Institute for Space Research, Landleven 12, 9747 AD Groningen, the Netherlands
           }

\date{Received / Accepted}
\abstract
{Observations of water lines are a sensitive probe of the geometry, dynamics and chemical structure of dense molecular gas. The launch of Herschel with on board HIFI and PACS allow to probe the behaviour of multiple water lines with unprecedented sensitivity and resolution.}
{We investigate the diagnostic value of specific water transitions in high-mass star-forming regions. As a test case, we apply our models to the AFGL\,2591 region.}
{A multi-zone escape probability method is used in two dimensions to calculate the radiative transfer. Similarities and differences of constant and jump abundance models are displayed, as well as when an outflow is incorporated.}
{In general, for models with a constant water abundance, the ground state lines, i.e., $\mathrm{1_{10}}$-$\mathrm{1_{01}}$, $\mathrm{1_{11}}$-$\mathrm{0_{00}}$, and $\mathrm{2_{12}}$-$\mathrm{1_{01}}$, are predicted in absorption, all the others in emission. This behaviour changes for models with a water abundance jump profile in that the line profiles for jumps by a factor of $\sim$\,10-100 are similar to the line shapes in the constant abundance models, whereas larger jumps lead to emission profiles. Asymmetric line profiles are found for models with a cavity outflow and depend on the inclination angle. Models with an outflow cavity are favoured to reproduce the SWAS observations of the $\mathrm{1_{10}}$-$\mathrm{1_{01}}$ ground-state transition. PACS spectra will tell us about the geometry of these regions, both through the continuum and through the lines.}
{ It is found that the low-lying transitions of water are sensitive to outflow features, and represent the excitation conditions in the outer regions. High-lying transitions are more sensitive to the adopted density and temperature distribution which probe the inner excitation conditions. The Herschel mission will thus be very helpful to constrain the physical and chemical structure of high-mass star-forming regions such as AFGL\,2591.}
\authorrunning{D. R. Poelman and F. F. S. van der Tak}
\titlerunning{Water line emission toward AFGL\,2591}
\maketitle

\section{\label{sec:intro}Introduction}
High-mass stars, more than their low-mass counterparts, play a critical role in the energetics and dynamical feedback into the interstellar medium (ISM). Despite this, the formation mechanisms of massive stars are still a puzzle, owing to their occurrence in more distant crowded stellar clusters, shorter formation timescales, and formation in regions of high visual extinction. Despite observational obstructions, promising theoretical work has been done on the formation of high-mass stars \cite[e.g.,][]{1998MNRAS.298...93B, 2001ASPC..243..139K, 2002Natur.416...59M, 2003ApJ...585..850M, 2007MNRAS.374L..29K}. Whereas low- and intermediate-mass stars ({\it M}\,$\lesssim$\,10\,M${_{\odot}}$) are formed through gravitational collapse and subsequent disk accretion \citep{1993ApJ...418..414P}, radiation pressure on the dust in the infalling gas prohibits high-mass stars to suffi\-ciently accrete gas and thus limits the stellar mass at 20\,--\,40\,M${_{\odot}}$ \cite[][]{1977A&A....54..183Y, 1987ApJ...319..850W, 1994ApJS...95..517B}. However, observations show that considerably more massive stars exist. Two formation scenarios have been proposed to work out this contradiction: non-spherical accretion \citep{2002ApJ...569..846Y} and merging of lower mass stars \citep{1998MNRAS.298...93B, 2002MNRAS.336..659B}. In order to distinguish between these two scenarios detailed observations, sensitive to a range of size scales and physical conditions, are required. For this, interferometers can provide the spatial distribution of dust and gas at the smallest size scales, while single dish telescopes equipped with large format, wide bandwidth, detectors are capable of simultaneously imaging a large number of molecules (albeit at lower spatial resolution). 

Up to date, there is growing evidence that disks are present around massive\- young stars as reports of direct imaging of disks in regions of high-mass star formation are increasing in number \cite[e.g.,][]{2004ApJ...601L.187B, 2002ApJ...566..982Z, 2003ApJ...590L..45S, 2005A&A...437..947V, 2006A&A...447.1011V}. In addition to direct imaging of disk-like structures, outflows provide indirect evidence for disks around massive young stars. In particular, collimated outflows are a sign of accretion \citep{2002A&A...383..892B}. Outflows are well known phenomena accompanied with sites of low-mass star formation, transporting angular momentum away from the star. The situation in the formation of high-mass stars is less clear, since spatial re\-solution has been lacking to resolve the outflows. However, over the last years, outflows from high-mass stars have received increasing attention \cite[see recent reviews by, e.g.,][]{2003ASPC..287..333S, 2005IAUS..227..237S, 2005Ap&SS.295....5C, 2007prpl.conf..245A}. It is found that the higher-mass objects appear to continue smoothly the corre\-la\-tion found in T Tauri stars between outflow and accretion signatures. These results suggest that the disk-outflow connection found in low-mass pre-main-sequence stars extends to more massive objects, and favour the accretion scenario. Besides this, \cite{2005ApJ...618L..33K} show that outflows around massive stars lead to a significant anisotropy in the stellar radiation field, hence reducing the radiation pressure experienced by the gas in the infalling envelope. As a result, a larger fraction of the material in the envelope can be used in the formation process of a massive star as opposed to models without outflows.

The dynamical and physical evolution during star formation is accompanied by strong chemical evolution. The cold-core phase, dominated by relatively simple molecular species, is transformed into a hotbed of more complex molecules after the protostellar object heats up the surrounding gas. Some of these molecules, e.g., H$_\mathrm{2}$, CO, and H$_\mathrm{2}$O, play an important role in regulating the temperature, hence pressure, via the process of heating and cooling through line absorption and emission. \cite{1997ApJ...489..122D} constructed models for the thermal balance, chemistry, continuum and line radiative transfer in dense molecular cloud cores with an embedded protostar. They find that due to temperature and density gradients, mo\-le\-cu\-lar abundances are not constant within dense molecular cloud cores, i.e., freeze-out of water onto dust grains for temperatures below the sublimation temperature of $\sim$\,100\,K and a high abundance of water in regions where the temperature is larger than 100\,K. Also, this behaviour has been found by \cite{2000A&A...355.1129C} who modeled the structure of the collapsing envelope around the low-mass protostar IRAS\,16293-2422. Besides this, outflows not only affect the physical state of the interstellar gas, but are also accompanied with chemical changes. Observational estimates of molecular abundances in outflow regions have suggested that particular species, e.g., SiO, CH$_\mathrm{3}$OH, HCN, H$_\mathrm{2}$CO, SO, and SO$_\mathrm{2}$ are significantly enhanced \cite[e.g.,][]{1997ApJ...487L..93B}. In particular for water, the elevated gas temperatures found in the outflows increase the rate coefficients for the neutral-neutral reactions, leading to an enhanced water production rate. Consequently, the abundance of water shows significant variations up to a few orders of magnitude from one region to another. This large variation in abundance makes water a powerful diagnostic of the physical structure of the region and of the fundamental chemical interactions between the gas and the grains. Observations, together with theoretical modeling of the available rotational and ro-vibrational lines from species and their isotopes, are neccesary to shed light on the density, temperature, and chemical structure over a wide range of conditions relevant for regions of star formation. Water in particular, with its many rotational lines, is very well suited to study the physical conditions in regions of low-mass \citep{vanKempen2007A&A} and high-mass (this paper) star formation.

In this paper, we model the excitation of water in the high-mass star-forming region AFGL\,2591. The aim of our research is to make predictions for the line strengths and profiles of various ortho- and para-H$_\mathrm{2}$O (o- and p-H$_\mathrm{2}$O) transitions to be observed with the {\it Heterodyne Instrument for the Far Infrared} (HIFI) and the {\it Photodetector Array Camera and Spectrometer} (PACS) on board of Herschel, and to study how the results depend on the adopted geometry and chemical structure. Resulting line strengths for specific transitions are then compared to existing {\it Infrared Space Observatory} (ISO) and {\it Submillimeter Wave Astronomy Satellite} (SWAS) observations.              

\section{The high-mass star-forming region AFGL\,2591}
While most massive stars form in clusters, GL\,2591, located in the Cygnus X region, provides one of the rare cases of a massive star forming in relative isolation. This allows to study the temperature, density, and velocity structure of the circumstellar envelope without confusion from nearby objects. Large columns of dust and gas toward GL\,2591 block our view of the stellar photosphere, making it invisible at optical wavelengths, but results in bright infrared emission. The infrared source is associated with a weak radio continuum source \citep{1984ApJ...287..334C, 2003ApJ...589..386T} and with a bipolar molecular outflow larger than 1$'$ in extent \citep{1984ApJ...286..302L, 1992ApJ...386..604M, 1995ApJ...451..225H}. Observations of the molecular cloud complexes in the Cygnus X region indicate that most objects are located at the same distance of 1.7\,kpc \citep{2006A&A...458..855S}. However, in the following we adopt a distance to the Cygnus X region of 1\,kpc, fixing the luminosity at $\sim$\,2\,$\times$\,$10^4$\,$\mathrm{L_{\odot}}$, to allow comparison with previous work \citep{1999ApJ...522..991V}. The results presented here depend weakly on distance, making the adopted distance inconsequential.  

AFGL\,2591 has been observed in water lines over a range of excitation conditions. \cite{1996A&A...315L.173H} report the detection of more than 30 lines within the bending vibration of water at 6\,$\mu$m using the {\em Short Wavelength Spectrometer} \citep[SWS,][]{1996A&A...315L..49D} on board ISO \citep[][]{1996A&A...315L..27K}. \cite{1996A&A...315L.177V} and \cite{2003A&A...403.1003B} derive, from observed $\mathrm{H_2O}$ spectra around 6\,$\mu$m, water abundances of $\sim$\,$10^{-5}$ up to $10^{-4}$ in hot-core regions. \cite{2000ApJ...539L.101S} observed with the {\em Submillimeter Wave Astronomy Satellite} (SWAS) the ground-state transition of o-$\mathrm{H_2O}$, thereby detecting the $\mathrm{H_2O}$ emission from the more extended, cold gas. They find that the $\mathrm{H_2O}$ abundances estimated for the hot core gas is at least 100 times larger than in the gas probed by SWAS, the latter being 6\,$\times$\,$10^{-10}$ to $10^{-8}$. \cite{2003A&A...406..937B} observed with the Short and Long Wavelength Spectrometer (SWS and LWS) on board of ISO and with SWAS $\mathrm{H_2O}$ lines toward deeply embedded massive protostars, and find that ice evaporation in the warmer envelope and freeze-out in the cold outer parts together with pure gas-phase chemistry reproduces the $\mathrm{H_2O}$ observations. This conclusion is strengthened by ground-based observations of the $\mathrm{H_2^{18}O}$ isotope \citep{2006A&A...447.1011V}.

Despite previous observations of the region, the future launch of ESA's Herschel Space Observatory \citep{2005dmu..conf....3P}, with on board the HIFI will provide key information on the physical and chemical conditions in molecular clouds by means of observing many water lines with higher sensitivity. Moreover, its high angular resolution allows to separate regions. HIFI, a heterodyne receiver, is designed to provide continuous frequency coverage from 480 to 1250\,GHz (Band 1\,--\,5), while Band 6 will cover the 1410\,--\,1910\,GHz frequency interval. Going up in frequency, the beam size decreases from 39$''$ to 13$''$, and together with a spectral resolution of 0.2\,--\,0.4\,km\,s$^{-\mathrm{1}}$ this instrument will be a powerful tool to probe the kinematics, i.e., infall and/or outflow, and chemical complexity of the AFGL\,2591 region. In addition, PACS will take spectral images over the spectral band from 57 to 210\,$\mu$m. The main advantages of this instrument are the 9$''$ resolution together with a 5\,$\times$\,5 pixel imaging capability. Although the PACS spectral resolution of $\lambda$/$\Delta\lambda$\,$\sim$\,1500 is insufficient to probe the kinematics in regions such as AFGL\,2591, mapping of the H$_\mathrm{2}$O and continuum emission will provide supplementary information over a wavelength range inaccessible to HIFI.

\section{Model set-up}
The temperature and density structure of the gas and dust in the envelope surrounding the young stellar object (YSO) AFGL\,2591 has been determined by \cite{1999ApJ...522..991V, 2000ApJ...537..283V}. In short, the resulting profiles are based on a study of single-dish submillimeter spectroscopy, combined with molecular line emission at (sub)-millimeter wavelengths. The temperature structure of the dust was modeled with the dust radiative transfer program of \cite{1988CoPhC..48..271E}, solving for the thermal balance of the grains as a function of distance to the star. The resulting temperatures follow a $r^{- 0.4}$ profile in the outer parts of the envelope, i.e., distances greater than 2--3\,$\times$\,$10^3$\,AU from the star, whereas in the inner envelope, the temperature gradient is steeper than $r^{- 0.4}$. The density structure is then obtained with the use of the RATRAN code \citep{2000A&A...362..697H} to model the observed molecular lines, i.e., CS, $\mathrm{HCO^+}$, HCN, and $\mathrm{H_2CO}$, using the temperature distribution calculated by the dust code. It is found that the density follows a power law of the form {\it n}\,=\,{\it n}$_\mathrm{0}$({\it r}/{\it r}$_\mathrm{0}$)$^{-\,\alpha}$, with $\alpha$\,$\sim$\,1\,--\,2.  Dust emission profiles are considered as well. In the following, results are presented for $\alpha$\,=\,1 and implications of an adopted $\alpha$\,=\,2 profile are elaborated on in Section \ref{sec:discussion}. 

The $\mathrm{H_2O}$ line emission in this work is calculated by application of the numerical code of \cite{2005A&A...440..559P, 2006A&A...453..615P} in which the radiative transfer is solved by means of a multi-zone escape probability method. The code, '$\beta$3D', is tested extensively against benchmark problems presented in \cite{2002A&A...395..373V} and \cite{2005dmu..conf..431V}. The radiative transfer in this work includes rotational transitions of the ground-vibrational state up to an upper energy level of $\sim$\,1000\,K. Vibrational pumping has been neglected as it is numerically challenging to include a substantial amount of ground-vibrational rotational levels, as well as levels of the vibrational mode in two dimensions. Spectroscopic and collisional input is taken from the molecular database as presented in \cite{2005A&A...432..369S}. Collisions between H$_\mathrm{2}$O and He, scaled to H$_\mathrm{2}$ to account for the different reduced mass of the collisional partner, are taken into account \citep{1993ApJS...85..181G}. The adopted gas and dust temperature in our models exceeds 20\,K at any point. Therefore, collisional rates presented in \cite{2006A&A...460..323D} for collisions of H$_\mathrm{2}$O with H$_\mathrm{2}$ in the temperature regime 5\,--\,20\,K are not taken into account. The influence of dust radiation on the excitation of the water molecule is taken into account using grain properties from \cite{1994A&A...291..943O}, Model 5.  We adopt the expression for the ortho-to-para ratio (OPR) of $\mathrm{H_2O}$, in thermal equilibrium, defined by 
\begin{equation}
\mathrm{OPR}= {{(2I_\mathrm{o} + 1)\sum(2J + 1)\exp\left(-{E_\mathrm{o}(J,K_\mathrm{a},K_\mathrm{c})\over kT}\right)}\over{(2I_\mathrm{p} + 1)\sum(2J + 1)\exp\left(-{E_\mathrm{p}(J,K_\mathrm{a},K_\mathrm{c})\over kT}\right)}}\,, 
\label{eq:OPR}
\end{equation}
where $I_\mathrm{o}$ and $I_\mathrm{p}$ are the total nuclear spin, corresponding to whether the hydrogen nuclear spins are parallel ($I_\mathrm{o}$\,=\,1, $\uparrow$$\uparrow$) or anti-parallel ($I_\mathrm{p}$\,=\,0, $\uparrow$$\downarrow$). The sum in the numerator (denominator) extends over all ortho (para) levels $({J},K_\mathrm{a},K_\mathrm{c})$; see \citet{1987A&A...187..419M}. 

\begin{figure}[t!]
\begin{minipage}[c]{0.5\textwidth}
\includegraphics[width=1\textwidth]{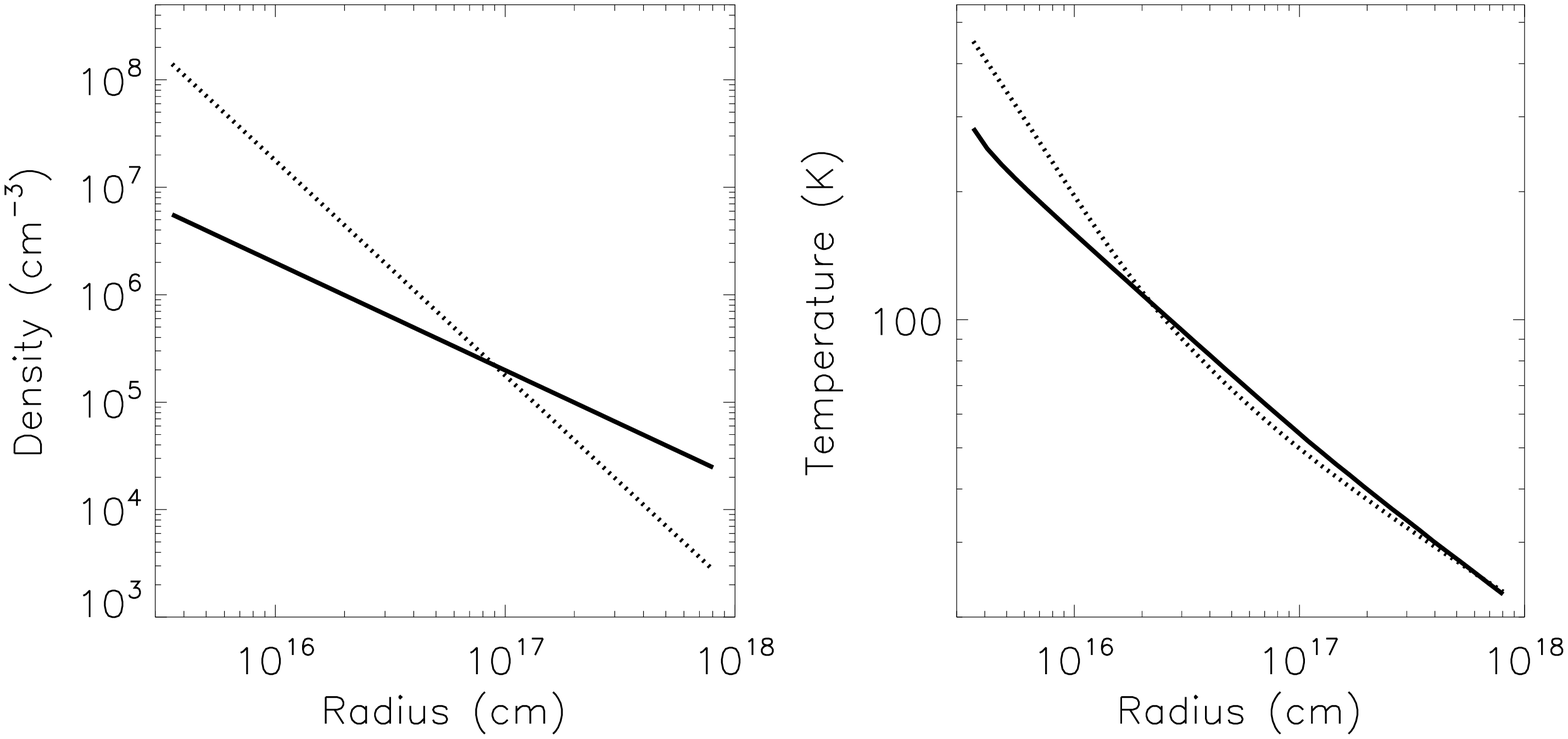}
\caption{The distribution of density ({\em left}) and temperature of gas and dust ({\em right}) as a function of radius in the cloud is plotted. Solid lines mean $\alpha$\,=\,1, dotted $\alpha$\,=\,2.}
\label{fig:density_temperature}
\end{minipage}
\end{figure}

Throughout this paper, a two-dimensional set-up is chosen in which the models consist of 78 logarithmically spaced shells with an outer radius of $\sim$\,50\,000\,AU. A constant Doppler parameter (1/e width) of 2\,km\,$\mathrm{s^{-1}}$ is assumed for all the models. Fig.\ \ref{fig:density_temperature} plots the adopted temperature and density structure as a function of radius for the models with $\alpha$\,=\,1 and 2. It is seen that the $\alpha$\,=\,1 models tend to have lower density and temperature towards the center of the cloud. The gas and dust tempe\-ra\-ture is assumed to be coupled, due to the high densities reached throughout most of the cloud. The calculated level populations at each position in the cloud are then used to compute velocity profiles of various transitions of o- and p-H$_\mathrm{2}$O using the program SKY that calculates the sky brightness distribution, part of the RATRAN code. SKY has been tested against other codes. For transitions with a rest frequency within the frequency coverage of HIFI, the resulting sky brightness distribution is convolved with the appropriate beam (depending on the frequency of the transition).  In addition, predictions are made for the spectrum to be observed with PACS. Line strengths and profiles are compared with existing ISO observations \citep{2003A&A...406..937B} and therefore convolved with the appropriate ISO-LWS beam, between 66$''$ and 78$''$ in diameter, depending on the frequency.

Due to the complex structure and spectroscopic properties of the water molecule, inverted transitions ({\em masers}) are likely to occur. Maser action, a process of intense radiation confined within small angular sizes with beaming angles in the range of $10^{-1}$\,--\,$10^{-2}$\,rad \citep{1992asma.book.....E}, necessitates higher angular resolution than achieved with our adopted multi-zone escape probability method. Moreover, a large negative optical depth, $\tau$, magnifies the exp(-$\tau$) term in the escape probability  which can hamper convergence. Therefore, we have artificially prevented the escape probabilities from exceeding exp(1), even when the level populations imply that -\,$\tau$\,$>$\,1. A maser transition becomes saturated in the regime that BJexp(-$\tau$)\,$\sim$\,$\Gamma$, with BJ the rate for stimulated emission and $\Gamma$ the effective pump rate. We have checked that this limit is not reached for this transition, or other maser lines. The masers produced by our models are thus unsaturated, although this conclusion depends on the details of the adopted velocity field. As a consequence, preventing the escape probabilities from exceeding exp(1) has no consequences on the calculated line intensities since non-saturated masers have negligible effect on the remaining level populations.  Nevertheless, because of this limitation no predictions are made for maser action, e.g., 22\,GHz maser transition. 

\section{Results}
\subsection{\label{subsec:constant}Constant water abundance}

\begin{figure}[t!]
\includegraphics[width=8.5cm]{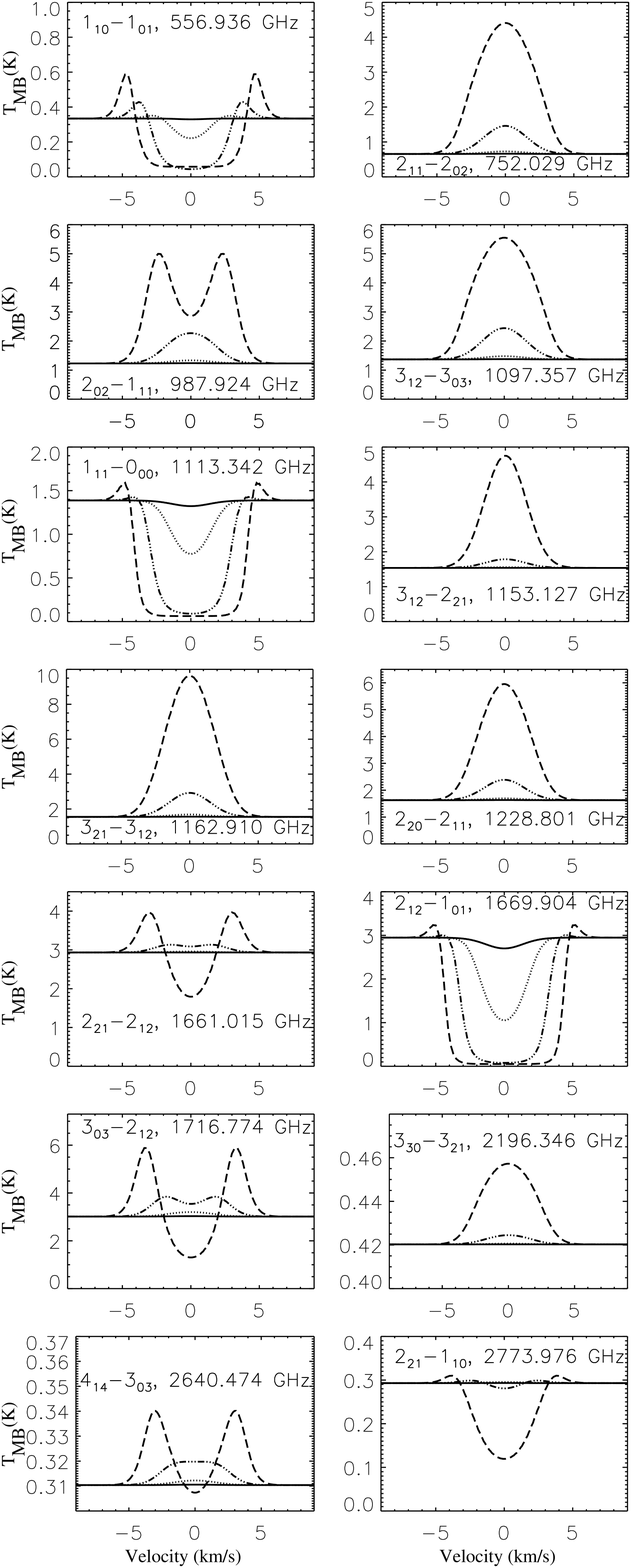}
\caption{Line profiles of ortho- and para-H$_\mathrm{2}$O transitions for models with $\alpha$\,=\,1. The water abundance, i.e., o-H$_\mathrm{2}$O\,+\,p-H$_\mathrm{2}$O, is uniformly distributed across the region and ranges from 10$^{-\mathrm{10}}$ to 10$^{-\mathrm{7}}$, with 10$^{-\mathrm{10}}$ ({\it solid}), 10$^{-\mathrm{9}}$ ({\it dotted}), 10$^{-\mathrm{8}}$ ({\it dash-dotted}), 10$^{-\mathrm{7}}$ ({\it dashed}).}
\label{fig:constant_alpha1}
\end{figure}

\begin{table}[b!]
\label{tab:constant_alpha1}
\centering
\begin{threeparttable}
\tabcolsep=1.6mm 
\caption{\small Integrated intensities, $\int${\it T}$_\mathrm{MB}$$\Delta${v} [K\,km\,s$^{-\mathrm{1}}$], for models with a constant H$_\mathrm{2}$O abundance.}
\centering
\begin{tabular}{llrrrr}
\hline 
\hline 
\multicolumn{1}{l}{Transition}&\multicolumn{1}{l}{Frequency}&\multicolumn{4}{c}{Water abundance}\\
&[GHz]&10$^{-\mathrm{10}}$&10$^{-\mathrm{9}}$&10$^{-\mathrm{8}}$&10$^{-\mathrm{7}}$\\
\hline
\multicolumn{6}{c}{ortho-H$_\mathrm{2}$O transitions}\\
\hline
1$_\mathrm{10}$-1$_\mathrm{01}$ & 557.936  &  $-$1.1($-$2) &  $-$0.2 &    $-$1.2 &  $-$1.4\\
3$_\mathrm{12}$-3$_\mathrm{03}$ & 1097.35 &   3.3($-$2) &   0.4 &     4.1 &  21.8\\
3$_\mathrm{12}$-2$_\mathrm{21}$ & 1153.12 &   6.0($-$3) &   7.2($-$2) & 0.9 &  12.0\\
3$_\mathrm{21}$-3$_\mathrm{12}$ & 1162.91 &   4.1($-$2) &   0.5 &     5.1 &  34.6\\
2$_\mathrm{21}$-2$_\mathrm{12}$ & 1661.01 &   2.0($-$3) &   0.1 &     1.1 &  0.6\\
2$_\mathrm{12}$-1$_\mathrm{01}$ & 1669.90 &  $-$0.8     &  $-$6.7 &   $-$17.7 &  $-$24.0\\
3$_\mathrm{03}$-2$_\mathrm{12}$ & 1716.77 &   6.4($-$2) &   0.7 &     4.6 &   5.0\\
3$_\mathrm{30}$-3$_\mathrm{21}$ & 2196.34 &   1.1($-$4) & 1.4($-$3)& 1.6($-$2)&  0.2\\
4$_\mathrm{14}$-3$_\mathrm{03}$ & 2640.47 &   6.6($-$4) &7.1($-$3)&  5.2($-$2)&   0.1\\
2$_\mathrm{21}$-1$_\mathrm{10}$ & 2773.97 &   7.9($-$4) &9.6($-$3)& $-$2.0($-$3)&$-$0.7\\
\hline
\multicolumn{6}{c}{para-H$_\mathrm{2}$O transitions}\\
\hline 
2$_\mathrm{11}$-2$_\mathrm{02}$ & 752.029  &   2.1($-$2)&   0.3 &  3.0 & 19.1\\
2$_\mathrm{02}$-1$_\mathrm{11}$ & 987.924  &   3.2($-$2)&   0.4 &  4.3 & 20.7\\
1$_\mathrm{11}$-0$_\mathrm{00}$ & 1113.34 &  $-$0.2    &  $-$2.0 &  $-$7.1 & $-$10.1\\
2$_\mathrm{20}$-2$_\mathrm{11}$ & 1228.80 &   2.0($-$2)&   0.2 &   2.8 & 18.7\\
\hline
\end{tabular}
N{\tiny OTE:} {\it a(b)} means {\it a}\,$\times$\,10$^\mathrm{\it b}$
\end{threeparttable}
\end{table}

As a first step, we assume the $\mathrm{H_2O}$ abundance to be uniformly distributed. \cite{2000ApJ...539L.101S} find, using SWAS observations of the 1$_\mathrm{10}$-1$_\mathrm{01}$ ground-state transition at 557\,GHz, that the abundance of o-H$_\mathrm{2}$O relative to H$_\mathrm{2}$ in giant molecular cloud cores varies between $\sim$\,10$^{-\mathrm{10}}$ and 10$^{-\mathrm{8}}$. In particular for AFGL\,2591, a value of 6\,$\times$\,10$^{-\mathrm{9}}$ is found. In view of this large observed spread, a constant H$_\mathrm{2}$O abundance with respect to H$_\mathrm{2}$, ranging from 10$^{-\mathrm{10}}$ to 10$^{-\mathrm{7}}$ is chosen. Resulting line profiles are plotted in Fig.\ \ref{fig:constant_alpha1}, integrated brightness temperatures are listed in Table \ref{tab:constant_alpha1}. Beside H$_\mathrm{2}^\mathrm{16}$O lines, multiple H$_\mathrm{2}^\mathrm{17}$O and H$_\mathrm{2}^\mathrm{18}$O transitions will be observed with HIFI and PACS for high-mass YSOs. Hence, model results in the case {\it X}(H$_\mathrm{2}$O) is 10$^{-\mathrm{10}}$ and 10$^{-\mathrm{9}}$ are representative for H$_\mathrm{2}^\mathrm{17}$O and H$_\mathrm{2}^\mathrm{18}$O, respectively. 

In general, the ground-state lines, i.e., 1$_\mathrm{10}$-1$_\mathrm{01}$, 1$_\mathrm{11}$-0$_\mathrm{00}$ and 2$_\mathrm{12}$-1$_\mathrm{01}$, appear in absorption. All other lines appear in emission. The optical depths in the 1$_\mathrm{10}$-1$_\mathrm{01}$, 1$_\mathrm{11}$-0$_\mathrm{00}$ and 2$_\mathrm{12}$-1$_\mathrm{01}$ transitions are factors of $\gtrsim$\,10 higher than the optical depths in the other lines, causing self-absorbed line profiles. When increasing the abundance by factors of 10, the line shapes transform from single-peaked emission into double-peaked self-absorbed profiles for the 2$_\mathrm{02}$-1$_\mathrm{11}$, 2$_\mathrm{21}$-2$_\mathrm{12}$, and 3$_\mathrm{03}$-2$_\mathrm{12}$ transitions. The strongest transitions are found to be those arising from the warm inner region, e.g., 3$_\mathrm{12}$-3$_\mathrm{03}$, 3$_\mathrm{21}$-3$_\mathrm{12}$, and 3$_\mathrm{03}$-2$_\mathrm{12}$.

\subsection{Water abundance jump profile}

\begin{figure}[h!]
\includegraphics[width=8.5cm]{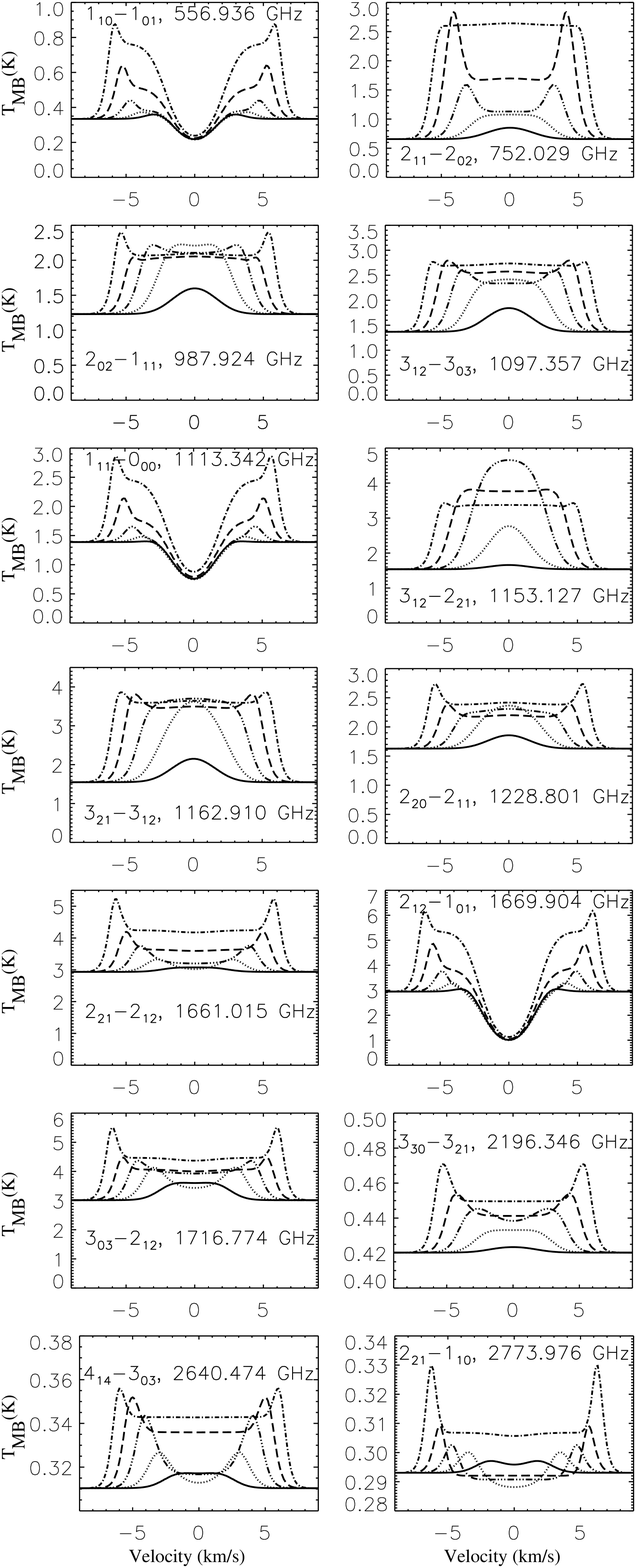}
\caption{Line profiles of H$_\mathrm{2}$O for the jump abundance models with $\alpha$\,=\,1. The water abundances for positions where {\it T}\,$<$\,100\,K is 10$^{-\mathrm{9}}$, whereas for {\it T}\,$>$\,100\,K the abundances range from 10$^{-\mathrm{8}}$ ({\it solid}), 10$^{-\mathrm{7}}$ ({\it dotted}), 10$^{-\mathrm{6}}$ ({\it dash dot dot}), 10$^{-\mathrm{5}}$ ({\it dashed}), 10$^{-\mathrm{4}}$ ({\it dash dot}).}
\label{fig:jump_alpha1}
\end{figure}

In this section we calculate the excitation of water, assuming that the $\mathrm{H_2O}$ distribution follows a step function, i.e., a low abundance in the outer regions and a higher abundance in the inner zones. \cite{2006A&A...454L...5D} have constructed models for the chemical evolution in a collapsing massive young stellar object (YSO). These models include a realistic evolution of the central source, as well as ad/desorption of ices from grain mantles as the grains fall in from the cool exterior into the warm interior. They find that this approach reproduces the step-function water distribution inferred observationally \citep{2006A&A...447.1011V}.  For this reason, models with a water abundance jump profile are computed. The position in the cloud where the jump occurs corresponds to the position where the sublimation temperature of water is reached, assumed to be 100\,K in our models. The water abundance outside the sublimation radius is kept constant at $10^{-9}$, whereas for regions with temperatures exceeding 100\,K, the water abundance is increased from $10^{-8}$ to $10^{-4}$ for the different models. Resulting line profiles are plotted in Fig. \ref{fig:jump_alpha1}, integrated brightness temperatures are listed in Table \ref{tab:jump_alpha1}. 

It is seen that the line profiles of the ground-state lines, i.e., 1$_\mathrm{10}$-1$_\mathrm{01}$, 2$_\mathrm{12}$-1$_\mathrm{01}$, and 1$_\mathrm{11}$-0$_\mathrm{00}$, are self-absorbed, similar to the constant abundance models. Line strengths and profiles of the ground-state transitions of o- and p-H$_\mathrm{2}$O, i.e., 1$_\mathrm{10}$-1$_\mathrm{01}$, 2$_\mathrm{12}$-1$_\mathrm{01}$ and 1$_\mathrm{11}$-0$_\mathrm{00}$, for models with a jump of a factor of 10\,--\,100 are equivalent to the line strengths and profiles in case of the model with a constant water abundance of 10$^{-\mathrm{9}}$. Larger jumps, however, do influence the line strengths and profiles for these transitions. Thus, the ground-state lines are partly influenced by the warm inner component, and partly by the cold outer layer. Higher-lying transitions differ from the constant abundance model in that larger line strengths are found even for moderate jumps. This is due to the enhanced inner warm water abundance, the region that favours the excitation of the high-lying transitions. 

\begin{table}[t!]
\label{tab:jump_alpha1}
\centering
\begin{threeparttable}
\tabcolsep=1.2mm 
\caption{\small Integrated intensities,$\int${\it T}$_\mathrm{MB}$$\Delta${v} [K\,km\,s$^{-\mathrm{1}}$], for models with a H$_\mathrm{2}$O abundance jump profile. The water abundance outside the sublimation radius is kept constant at 10$^{-\mathrm{9}}$.}
\centering
\begin{tabular}{llrrrrr}
\hline 
\hline 
\multicolumn{1}{l}{Transition}&\multicolumn{1}{l}{Frequency}&\multicolumn{5}{c}{Inner water abundance}\\
&[GHz]&10$^{-\mathrm{8}}$&10$^{-\mathrm{7}}$&10$^{-\mathrm{6}}$&10$^{-\mathrm{5}}$&10$^{-\mathrm{4}}$\\
\hline
\multicolumn{7}{c}{ortho-H$_\mathrm{2}$O transitions}\\
\hline
1$_\mathrm{10}$-1$_\mathrm{01}$ & 557.936  &  $-$0.2    &  $-$0.2    & 0.1     & 1.2    & 3.5 \\
3$_\mathrm{12}$-3$_\mathrm{03}$ & 1097.35 &   1.9    &  6.4     & 9.6     & 13.2   & 17.0 \\
3$_\mathrm{12}$-2$_\mathrm{21}$ & 1153.12 &   0.4    &  4.5     & 17.3    & 19.8   & 20.5\\
3$_\mathrm{21}$-3$_\mathrm{12}$ & 1162.91 &   2.3    &  9.9     & 17.0    & 21.2   & 25.9 \\
2$_\mathrm{21}$-2$_\mathrm{12}$ & 1661.01 &   0.7    &  2.0     &  4.6    & 9.3    & 18.7 \\
2$_\mathrm{12}$-1$_\mathrm{01}$ & 1669.90 &  $-$6.6   &  $-$6.3    &  $-$4.5   & 1.7    & 15.3 \\
3$_\mathrm{03}$-2$_\mathrm{12}$ & 1716.77 &   3.3    &   6.3    &   10.8  & 13.2   & 21.1 \\
3$_\mathrm{30}$-3$_\mathrm{21}$ & 2196.34 &   1.3($-$2)&   7.7($-$2)&   0.2   & 0.3    & 0.4\\
4$_\mathrm{14}$-3$_\mathrm{03}$ & 2640.47 &   3.9($-$2)&   8.1($-$2)&  0.2    & 0.3    & 0.5\\ 
2$_\mathrm{21}$-1$_\mathrm{10}$ & 2773.76  &   2.3($-$2)&   6.0($-$3)&  9.0($-$3)& 2.7($-$2)& 0.2\\
\hline
\multicolumn{7}{c}{para-H$_\mathrm{2}$O transitions}\\
\hline 
2$_\mathrm{11}$-2$_\mathrm{02}$ & 752.029  & 0.8  & 2.5  & 5.5   & 13.4 & 21.3\\
2$_\mathrm{02}$-1$_\mathrm{11}$ & 987.924  & 1.4  & 5.7  & 7.8   & 8.6  & 12.5\\
1$_\mathrm{11}$-0$_\mathrm{00}$ & 1113.34 &$-$1.9  &$-$1.9  &$-$1.2   & 1.5  & 7.4\\
2$_\mathrm{20}$-2$_\mathrm{11}$ & 1228.80 & 0.8  & 3.5  & 5.3   & 6.5  & 11.2\\
\hline
\end{tabular}
N{\tiny OTE:} {\it a(b)} means {\it a}\,$\times$\,10$^\mathrm{\it b}$
\end{threeparttable}
\end{table}

\begin{figure}
\includegraphics[width=0.5\textwidth]{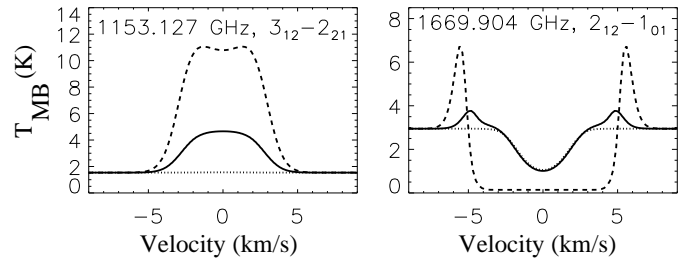}
\caption{Line profiles of two ortho-$\mathrm{H_2O}$ transitions. The solid line represents the line profile for a jump with an abundance of $10^{-9}$ where T\,$<$\,100\,K and $10^{-6}$ where T\,$>$\,100\,K. The dotted and dashed line profiles represent the constant water abundance of $10^{-9}$ and $10^{-6}$, respectively.}
\label{fig:jumpANDconstant}
\end{figure}

The peculiar appearance of the line profiles can be explained as a super\-position of two constant models, i.e., a small, warm and abundant component together with a cold, large and low abundant component. In other words, a peak or drop in the line profile corresponds to a peak or drop in one of the transitions, when the contributing lines are optically thin. However, the line profile when one or both of the transitions are optically thick is not a simple superposition. In this case, photons coming from line center are lost and only those photons that escape in the line wings will contribute to the final line profile. This can be seen in Fig.\ \ref{fig:jumpANDconstant} where we plot, for two different transitions, the resulting jump profiles with an inner water abundance of 10$^{-\mathrm{6}}$, together with the constant water abundance profiles of 10$^{-\mathrm{9}}$ and 10$^{-\mathrm{6}}$. It is seen that for the 2$_\mathrm{12}$-1$_\mathrm{01}$ transition, the contribution from the optically thick inner 10$^{-\mathrm{6}}$ part is limited to those photons escaping in the line wings. However, the resulting jump profile is dominated by the optically thin outer 10$^{-\mathrm{9}}$ component. On the contrary, the jump profile for the 3$_\mathrm{12}$-2$_\mathrm{21}$ transition is most heavily influenced by the inner component.

One can wonder whether HIFI's future high spectral resolution observations can distinguish between a constant or a jump profile water distribution. Since high-lying transitions, e.g., 3$_\mathrm{12}$-3$_\mathrm{03}$, 3$_\mathrm{21}$-3$_\mathrm{12}$, 2$_\mathrm{21}$-2$_\mathrm{12}$, more than the ground-state lines, are influenced by the warm inner water abundance, these lines are favoured to examine the occurence of an abundance jump. HIFI's high sensitivity allows to probe features in the line profile of $\sim$\,1\,K in Band 6, down to $\sim$\,0.05\,--\,0.1\,K in Band 1. This 5$\sigma$ detection is reached with an integration time of 1 hour for R\,=\,10$^\mathrm{4}$. Despite the differences in line strengths, a jump of a factor of $\sim$\,10\,--\,100 leads to line shapes similar to the profiles in the constant abundance models. Hence, we do not expect HIFI to be able to distinguish between the two scenarios for jumps up to a factor of $\sim$\,100. However, jumps of a factor of $\gtrsim$\,1000 lead to line profiles more flattened around line center, with peaks in the line wings. This feature is not seen in the constant models for abundances up to 10$^{-\mathrm{7}}$ for the high-lying transitions, and thus can be used to determine the water abundance structure. Nonetheless, to distinguish between constant and jump water abundance models is not as straightforward. It is likely that line profiles change under the influence of inhomogeneity and velocity gradients \citep{2006A&A...453..615P}. Hence, one can easily misinterpret the observed line profiles. Multi-line observations in the future will help to further disentangle the events that are responsible for the observed line profile.

\subsection{\label{subsec:cavity}Cavity}

\begin{figure}
\includegraphics[width=8.5cm]{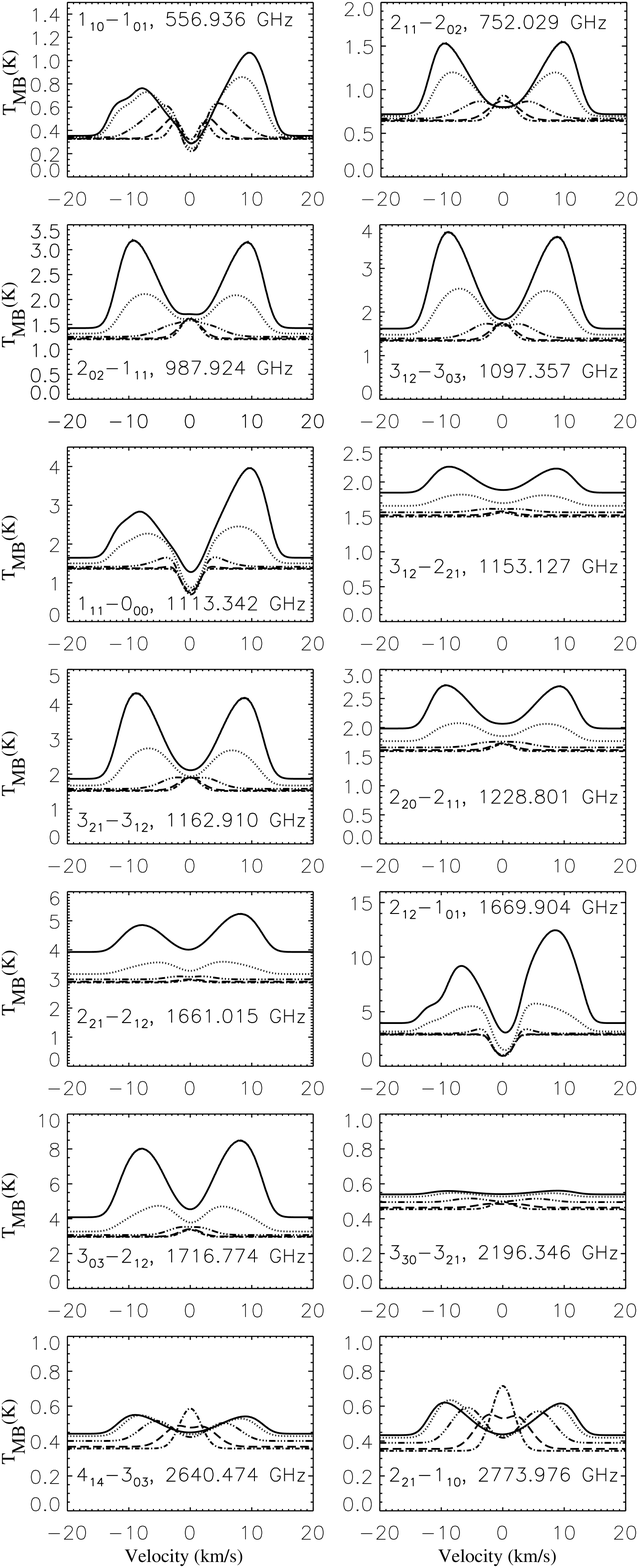}
\caption{Line profiles of H$_\mathrm{2}$O for a model with $\alpha$\,=\,1, a cavity with half-opening angle $\theta$\,=\,15$^{\circ}$ and an outflow velocity field with {\it V}$_\mathrm{out}$\,=\,12\,km\,s$^{-\mathrm{1}}$. The abundance outside the cavity has a constant value of 10$^{-\mathrm{9}}$, whereas inside the cavity the abundance is 10$^{-\mathrm{7}}$. The different line profiles are the result of the different inclination angles. Inclination of 0$^{\circ}$ means face-on, as the line of sight is directed along the cavity; inclination of 90$^{\circ}$ is edge-on. Inclination\,=\,0$^{\circ}$ ({\it solid}), 20$^{\circ}$ ({\it dotted}), 45$^{\circ}$ ({\it dash dot dot}), 70$^{\circ}$ ({\it dashed}), 90$^{\circ}$ ({\it dash dot}).}
\label{fig:cavity_alpha1}
\end{figure}

\begin{table}[!b]
\label{tab:cavity_alpha1}
\centering
\begin{threeparttable}
\caption{\small Integrated intensities,$\int${\it T}$_\mathrm{MB}$$\Delta${v} [K\,km\,s$^{-\mathrm{1}}$], for a model with an outflow cavity with half-opening angle $\theta$\,=\,15$^{\circ}$.}
\centering
\begin{tabular}{llrrrrr}
\hline 
\hline 
\multicolumn{1}{l}{Transition}&\multicolumn{1}{l}{Frequency}&\multicolumn{5}{c}{Inclination angle}\\
&[GHz]&0$^{\circ}$&20$^{\circ}$&45$^{\circ}$&70$^{\circ}$&90$^{\circ}$\\
\hline
\multicolumn{7}{c}{ortho-H$_\mathrm{2}$O transitions}\\
\hline
1$_\mathrm{10}$-1$_\mathrm{01}$ & 557.936  & 8.7   & 6.8   & 3.1  & 1.1      & 0.7\\
3$_\mathrm{12}$-3$_\mathrm{03}$ & 1097.35 & 32.9  & 17.4  & 3.7  & 1.8      & 1.6\\
3$_\mathrm{12}$-2$_\mathrm{21}$ & 1153.12 & 5.5   & 2.7   & 0.5  & 0.3      & 0.2\\
3$_\mathrm{21}$-3$_\mathrm{12}$ & 1162.91 &  36.6 & 17.8  & 3.4  & 1.7      & 1.5\\
2$_\mathrm{21}$-2$_\mathrm{12}$ & 1661.01 &  17.5 & 7.0   & 0.8  & 0.4      & 0.3\\
2$_\mathrm{12}$-1$_\mathrm{01}$ & 1669.90 &  105.8& 33.9  & $-$5.0 & $-$7.2     & $-$6.8\\
3$_\mathrm{03}$-2$_\mathrm{12}$ & 1716.17 &  67.2 & 25.3  & 3.6  &  1.8     & 1.5\\
3$_\mathrm{30}$-3$_\mathrm{21}$ & 2196.34 &  0.3  & 0.3   & 0.3  &  0.2     & 0.2\\  
4$_\mathrm{14}$-3$_\mathrm{03}$ & 2640.47 &  1.6  & 1.6   & 1.4  &  1.1     & 1.0\\
2$_\mathrm{21}$-1$_\mathrm{10}$ & 2773.97 &  2.7  & 2.8   & 2.5  &  1.9     & 1.7\\
\hline
\multicolumn{7}{c}{para-H$_\mathrm{2}$O transitions}\\
\hline 
2$_\mathrm{11}$-2$_\mathrm{02}$ & 752.029  & 12.1  &  8.1  & 2.7  & 1.3    & 1.2\\
2$_\mathrm{02}$-1$_\mathrm{11}$ & 987.924  & 26.6  &  13.7 & 3.1  & 1.7    & 1.6\\
1$_\mathrm{11}$-0$_\mathrm{00}$ & 1113.34 & 25.6  &  12.2 & $-$0.3 &$-$1.8    &$-$1.7\\
2$_\mathrm{20}$-2$_\mathrm{11}$ & 1228.80 & 11.2  &  5.3  & 1.0  & 0.5    & 0.5\\
\hline
\end{tabular}
\end{threeparttable}
\end{table}

We next consider cylindrically symmetric models in which an outflow cavity is incorporated. Observations confirm a powerful bipolar molecular outflow associated with AFGL\,2591 \citep[e.g.,][]{1983ApJ...265..824B, 1984ApJ...286..302L}, $\sim$\,1.5\,pc in extent. In addition, \cite{2006A&A...447.1011V} constructed a representation of the geometry of AFGL\,2591, based on millimeter interferometric observations. They conclude that they are observing a young protostar with a disk close to face-on (inclination between 26--38 degrees). Therefore, models are constructed to study the effect on the line strengths and profiles of a low-density outflow with an opening half-angle $\theta$, as function of inclination angle. For this, a model with an outflow cavity with half-opening angle $\theta$\,=\,$15^{\circ}$ is computed. Models with larger opening angles are also computed and discussed in Section \ref{sec:discussion}. The water abundance outside the cone is set at a constant value of $10^{-9}$, and the density and temperature structure follow Fig. \ref{fig:density_temperature}. The outflow is characterised as a low density region with n($\mathrm{H_2}$)\,=\,5\,$\times$\,$10^3$\,$\mathrm{cm^{-3}}$, in agreement with the overall density found through hydrodynamical modeling of outflows \citep{2006ApJ...649..845S}, and a coupled gas and dust temperature of 300\,K. This high gas and dust temperature is reasonable as in outflows the temperatures reach a few 100\,K \citep{2001ApJ...555...40G}. Due to these high temperatures, enhanced neutral-neutral rates lead to elevated $\mathrm{H_2O}$ abundances. For this reason, a constant water abundance of $10^{-7}$ is chosen throughout the outflow. The outcome of models with a water abundance of $10^{-6}$ in the outflow is discussed in Section \ref{sec:discussion}. The outflow velocity field is modeled as a function of radius, i.e., V\,$\propto$\,R, reaching a maximum velocity of $\pm$\,12\,km\,$\mathrm{s^{-1}}$ at the outer radius. Resulting line profiles are plotted in Fig. \ref{fig:cavity_alpha1}, integrated line strengths are listed in Table \ref{tab:cavity_alpha1}. 

One notices the difference in width of the line profiles for models with an inclination angle of $0^{\circ}$, compared to $90^{\circ}$. This is the result of the fraction of the outflow velocity field lying in the line of sight, i.e., face-on models tend to have a larger line profile width as opposed to edge-on models. It is seen from Table \ref{tab:cavity_alpha1} that a decrease in inclination angle leads to an increase in the line strengths as the cavity is more uncovered. The outflow is best seen in the ground-state transitions for models with a small inclination angle, i.e., close to face-on. This is expected since the low density cone favours the emission from low-lying transitions, in that the critical densities of the water lines are of the order of $10^{8}$\,$\mathrm{cm^{-3}}$. From Fig. \ref{fig:cavity_alpha1} we see that the peak in the blue wing is weaker than the peak in the red wing of the line profile for the $\mathrm{1_{10}}$-$\mathrm{1_{01}}$ ground-state transition, in contrast with the line profiles of the constant abundance and jump profile models. Note that for an inclination angle of $0^{\circ}$, the optical depth through the center, i.e., along the outflow, is larger than 1 for ground-state transition. It is found that models with an inclination angle of $90^{\circ}$ produce line profiles similar to the profiles in the case of a model with a constant water abundance of $10^{-9}$, except for the transitions seen with ISO. However, line strengths differ by a factor of a few.

Clearly, HIFI's sensitivity can put constraints on the inclination angle of the outflow, in that for small inclination angles, the blue wing is found to be less strong than the red wing.

\subsection{Disk}
Even though the formation of high-mass stars is not completely understood, more and more observational evidence is found that the creation originates in a similar manner as its low-mass counterpart \citep{1999A&A...345..949C}. Therefore, we examine the influence of an embedded disk in the center of the AFGL\,2591 region on the line profiles and strengths. We consider the scenario of Section \ref{subsec:constant}, for which we adopt a constant water abundance of $10^{-9}$ throughout the region.

The size of a circumstellar disk can be quantified as {\it r}$_\mathrm{d}$\,=\,350\,AU ($\gamma$/0.007) ($\mathrm{\it M}$${_{\star}}$/20\,M${_{\odot}}$)$^{\mathrm{1/2}}$, where $\gamma$ is the ratio of rotational to gravitational energy. For high-mass star-formation, $\gamma$\,$\sim$\,0.007 \citep{2003A&A...405..639P}. When adopting a mass of the central star toward AFGL\,2591 of 16\,$\mathrm{M_{\odot}}$, as inferred by \cite{2005A&A...437..947V}, a disk with radius $\sim$\,300\,AU is found. We adopt an overall disk density of 10$^\mathrm{9}$\,cm$^{-\mathrm{3}}$, and a water abundance of 10$^{-\mathrm{6}}$. The temperature of the gas and dust is assumed to be 100\,K. Note that the characterization of the density and temperature structure of the disk is far from accurate. However, the intent of this section is to investigate the influence of a 'simple' disk model, not to model a realistic disk. The excitation of rotational, as well as ro-vibrational, water transitions in a circumstellar disk when the chemistry and temperature structure is calculated self-consistently is under construction \citep{2007A&A}.

It is found that the line profiles and strengths are insignificantly influenced under all possible inclination angles (less than 1\,$\%$). Since HIFI will be able to spectrally resolve disk structures, the emission from our adopted disk model with a diameter up to a few hundred AU in size is overwhelmed by that of the envelope within the HIFI beams. However, observations show that more massive, elongated disks with a diameter up to few thousands of AU are likely to exist \citep{2005A&A...435..901B}. Due to high densities and large columns of dust in circumstellar disks, the stellar radiation field is attenuated. As a result the temperature of the gas and dust at radial distances of a more than a few 10 of AU drops below 100\,K \cite[see][]{2007ApJ...656..515G}. Consequently, water will be frozen out on dust grains, reducing significantly the gas phase water abundance. When adopting the same disk density as in previous model, a water abundance value of $10^{-8}$, and a disk radius of 1500\,AU, the line profiles differ only a few $\%$ from the models without a disk.

\subsection{Predictions for PACS}

\begin{figure}[h!]
\includegraphics[width=8.5cm]{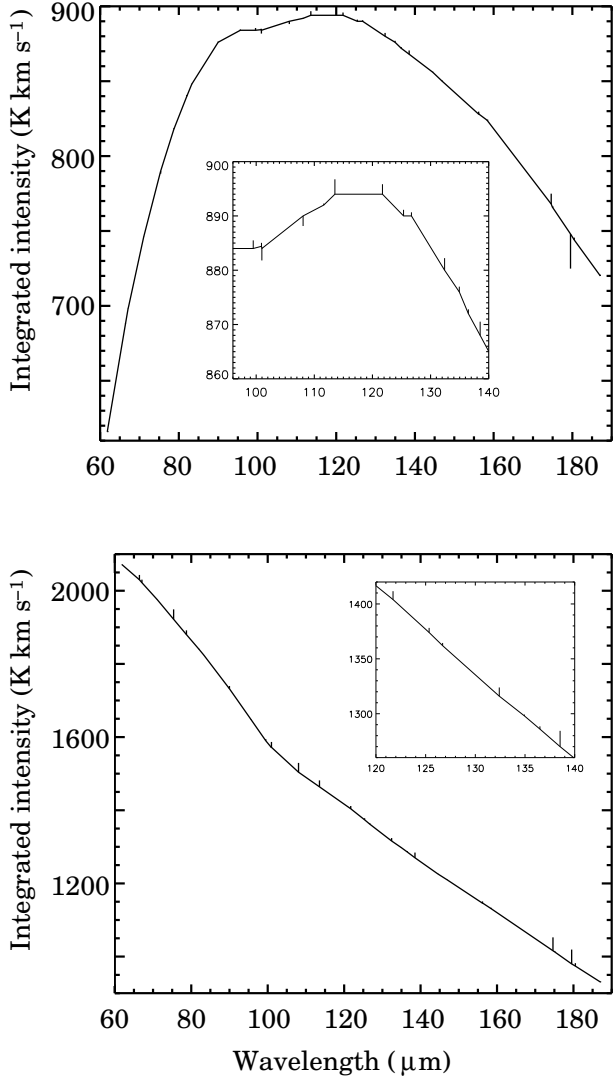}
\caption{Predictions for the spectrum to be observed with PACS for two different scenarios. Integrated intensities are plotted on top of the continuum background. The top panel denotes the spectrum for a model with a constant water abundance of 10$^{-\mathrm{8}}$. Bottom panel plots the spectrum when a cavity with an outflow velocity field, with {\it V}$_\mathrm{out}$\,=\,12\,km\,s$^{-\mathrm{1}}$, is seen under an inclination angle of 0$^{\circ}$. The Y-axis is in K\,km\,s$^{-\mathrm{1}}$. To convert to W\,cm$^{-\mathrm{2}}$, multiply by $\nu$$^\mathrm{3}$\,$\times$\,10$^{-\mathrm{57}}$.}
\label{fig:PACS}
\end{figure}

\begin{table}[t!]
\label{tab:PACS}
\centering
\begin{threeparttable}
\caption{\small Integrated intensities,$\int${\it T}$_\mathrm{MB}$$\Delta${v} [K\,km\,s$^{-\mathrm{1}}$], for two different models, when convolved with the PACS beam.}\centering
\begin{tabular}{llrr}
\hline 
\hline 
Transition&Wavelength& Model\,I & Model\,II\\
&[$\mu$m]&&\\
\hline
\multicolumn{4}{c}{ortho-H$_\mathrm{2}$O transitions}\\
\hline
3$_\mathrm{30}$-2$_\mathrm{21}$ & 66.48  & 0.3    &  13.2  \\ 
3$_\mathrm{21}$-2$_\mathrm{12}$ & 75.43  & $-$1.7 &  27.0  \\ 
4$_\mathrm{23}$-3$_\mathrm{12}$ & 78.79  & 0.6    &  11.7  \\ 
5$_\mathrm{05}$-4$_\mathrm{14}$ & 99.56  & 1.5    &  4.3   \\ 
2$_\mathrm{21}$-1$_\mathrm{10}$ & 108.1 & $-$1.9 &  25.4   \\ 
4$_\mathrm{14}$-3$_\mathrm{03}$ & 113.5 & 2.7    &  17.6   \\ 
4$_\mathrm{32}$-4$_\mathrm{23}$ & 121.7 & 1.8    &  7.6    \\ 
4$_\mathrm{23}$-4$_\mathrm{14}$ & 132.4 & 2.2    &  7.9    \\ 
3$_\mathrm{03}$-2$_\mathrm{12}$ & 174.6 & 6.9    &  36.5   \\ 
2$_\mathrm{12}$-1$_\mathrm{01}$ & 179.5 & $-$23.1  &39.2   \\ 
\hline
\multicolumn{4}{c}{para-H$_\mathrm{2}$O transitions}       \\
\hline 
3$_\mathrm{31}$-2$_\mathrm{20}$ & 67.13  & 0.2   &  8.3    \\ 
3$_\mathrm{22}$-2$_\mathrm{11}$ & 89.99  & 0.4   &  7.6    \\ 
2$_\mathrm{20}$-1$_\mathrm{11}$ & 100.9 & $-$2.2  & 14.8   \\ 
3$_\mathrm{13}$-2$_\mathrm{02}$ & 138.5 & 2.5   &  14.4    \\ 
\hline
\end{tabular}
N{\tiny OTE:} In Model\,I a constant water abundance of 10$^{-\mathrm{8}}$ is assumed. Model\,II describes AFGL\,2591 as a region with a cavity outflow, with {\it V}$_\mathrm{out}$\,=\,12\,km\,s$^{-\mathrm{1}}$, seen face-on, i.e., inclination angle\,=\,0$^{\circ}$. To convert to W\,cm$^{-\mathrm{2}}$, multiply $\int$$\mathrm{\it T}$$_\mathrm{MB}$$\Delta$$\mathrm{v}$ by $\nu$$^3$\,$\times$\,10$^{-57}$.
\end{threeparttable}
\end{table}

Besides HIFI detections, observations by PACS will enlarge significantly our knowledge on high-mass star-forming regions. PACS will carry out spectral scans from 57 to 210\,$\mu$m, thereby providing information on the dust continuum emission at wavelengths currently poorly characterized. However, the PACS data will be spectrally unresolved ($\lambda$/$\Delta\lambda$\,$\sim$\,1500), in contrast to HIFI observations. 

In this section, we compute the spectral output for PACS for two diffe\-rent scenarios. The first scenario (Model\,I) is for a model with a constant water abundance of 10$^{-\mathrm{8}}$, the second scenario (Model\,II) is for the model as described in Section \ref{subsec:cavity}, in which the outflow cavity is seen under an inclination angle of 0$^{\circ}$. Resulting spectra are plotted in Fig.\,\ref{fig:PACS}, integrated line strengths are listed in Table \ref{tab:PACS}. 

It is seen from Fig.\,\ref{fig:PACS} that the two models lead to rather different output spectra. Strong lines are 2$_\mathrm{20}$-1$_\mathrm{11}$, 2$_{\mathrm21}$-1$_\mathrm{10}$, 4$_\mathrm{14}$-3$_\mathrm{03}$, 3$_\mathrm{03}$-2$_\mathrm{12}$, and 2$_\mathrm{12}$-1$_\mathrm{01}$. Depending on the model, the 3$_\mathrm{21}$-2$_\mathrm{12}$, 2$_\mathrm{20}$-1$_\mathrm{11}$, 2$_\mathrm{21}$-1$_\mathrm{10}$, and 2$_\mathrm{12}$-1$_\mathrm{01}$ transitions are found either in emission or in absorption. One notices the change in the behaviour of the dust continuum emission between the two models. Hence, PACS spectra will tell us about the geometry of these regions, both through the continuum and through the lines. Note that the signal-to-noise (S/N) ratios for the different lines vary between 30 and 3000 for Model\,I and between 25 and 300 for Model\,II.

It is now interesting to compare the results from \cite{1996ApJ...456..611K} in which far-infrared water emission from shock waves is presented. They find that the strongest transitions are the 2$_{12}$-1$_{10}$, 3$_{03}$-2$_{12}$, 2$_{21}$-1$_{10}$, 4$_{14}$-3$_{03}$, 3$_{13}$-2$_{02}$, 1$_{10}$-1$_{01}$, and 3$_{22}$-2$_{11}$ lines. We find that these transitions are also the strongest in the case of models with an outflow cavity. 

\section{\label{sec:discussion}Discussion}

\subsection{Comparison to SWAS and ISO observations}

Observations by SWAS of the ground-state transition of o-$\mathrm{H_2O}$ at 557\,GHz reveal an asymmetric line profile \citep{2003A&A...406..937B}. However, radiative transfer modeling of line profiles in the massive young stellar object AFGL\,2591 show that constant and jump water profile distributions do not fit this SWAS observation of the ground-state transition of o-H$_\mathrm{2}$O, in that a symmetric line profile is found. In contrast to these models, an asymmetric line profile is derived for the ground-state transition for models with an outflow cavity and an inclination angle between 0$^{\circ}$ and 30$^{\circ}$. This agrees with the observations by \cite{2006A&A...447.1011V}, who conclude that the outflow is seen under an inclination angle of 26$^{\circ}$--\,38$^{\circ}$. The integrated intensity of $\pm$\,7\,--\,9\,K\,km\,s$^{-\mathrm{1}}$ reduces, after convolution with a beam size comparable to the size of the SWAS beam, i.e.,  3.3$'$\,$\times$\,4.5$'$, to a value of $\pm$\,0.5\,K\,km\,s$^{-\mathrm{1}}$. This value is in agreement with the observed 0.37\,$\pm$\,0.04\,K\,km\,s$^{-\mathrm{1}}$ strength. Note that this observed value is a lower limit, and that the actual value may be increased by a factor of $\sim$\,2, due to the presence of foreground clouds as discussed in \cite{2003A&A...406..937B}. 

\begin{table}[t!]
\label{tab:ISOANDSWAS}
\centering
\begin{threeparttable}
\caption{\small Observed ISO and SWAS line intensities [\mbox{K\,km\,s$^{-\mathrm{1}}$}] for AFGL\,2591, see Boonman et al.\ 2003}
\small
\centering
\begin{tabular}{cccc}
\hline 
\hline 
\multicolumn{4}{c}{Transition and Wavelength [GHz]}\\
\hline
2773.97&2640.47&2196.34&557.936 \\
2$_\mathrm{21}$-1$_\mathrm{10}$&4$_\mathrm{14}$-3$_\mathrm{03}$ & 3$_\mathrm{30}$-3$_\mathrm{21}$& 1$_\mathrm{10}$-1$_\mathrm{01}$ \\
\hline
$<$\,$-$0.5&$<$\,0.7&$<$\,1.2&0.37\,$\pm$\,0.04\\
\hline
\end{tabular}
\end{threeparttable}
\end{table}

Table \ref{tab:ISOANDSWAS} lists the observed integrated line intensities for some selected ISO transitions, published in \cite{2003A&A...406..937B}.  The observed line fluxes [W\,cm$^{-\mathrm{2}}$\,$\mu$m$^{-\mathrm{1}}$] have been converted into integrated line intensities [K\,km\,s$^{-\mathrm{1}}$], by adopting the appropriate ISO beam size, to simplify comparison with the results presented in this chapter. For the 3$_\mathrm{30}$-3$_\mathrm{21}$ and 4$_\mathrm{14}$-3$_\mathrm{03}$ transitions, the predicted integrated intensities are below the observed strengths for all the models by a factor of $\sim$\,4, except for the models with an outflow that overproduce the 4$_\mathrm{14}$-3$_\mathrm{03}$ strength by a factor of $\sim$\,2. The 2$_\mathrm{21}$-1$_\mathrm{10}$ line is observed in absorption, whereas our models favour to produce this transition in emission. The model with a constant water abundance of 10$^{-\mathrm{7}}$ is favoured to fit this transition. 

Modeling efforts of this paper are similar to the results presented in \cite{2003A&A...406..937B} in that {\it (i)} the 4$_\mathrm{14}$-3$_\mathrm{03}$, 3$_\mathrm{30}$-3$_\mathrm{21}$ and 3$_\mathrm{03}$-2$_\mathrm{12}$ transitions are always predicted to be in emission for models with $\alpha$\,=\,1. {\it (ii)} The 2$_\mathrm{21}$-1$_\mathrm{10}$ and 2$_\mathrm{12}$-1$_\mathrm{01}$ lines are found in emission for models with an outer abundance of 10$^{-\mathrm{9}}$. {\it (iii)} Even though we tabulate in Section \ref{subsec:constant} the resulting integrated intensities up to a constant water abundance of 10$^{-\mathrm{7}}$, the trend is visible that models with a larger abundance would result in a deeper absorption of the 2$_\mathrm{21}$-1$_\mathrm{10}$ transition than observed. Hence, models without freeze-out are not favoured to model this transition. {\it (iv)} \cite{2003A&A...406..937B} conclude that models with an inner water abundance of 2\,$\times$\,10$^{-\mathrm{4}}$ and an outer water abundance in between 10$^{-\mathrm{11}}$ and 10$^{-\mathrm{8}}$ are favoured the match the observations for all the transitions. Our jump model with an inner water abundance of 10$^{-\mathrm{4}}$ produces integrated intensities that match closely these preferred models.

However, due to the lack of more o- and p-H$_\mathrm{2}$O observations and the poor spectral resolution of the ISO satellite, the ISO-LWS data are not decisive on which model is favoured. The Herschel mission will bring additional (line profile) information on water transitions with somewhat lower energies, which is needed to decide between the different models. Therefore, preparation work, as presented in this paper and \cite{2005prpl.conf.8537D}, are valuable to prepare for future HIFI observing programs of high-mass star-forming regions.

\subsection{Parameter dependency}
\subsubsection{Steepness of density gradient}

\begin{figure}[h!]
\includegraphics[width=8.5cm]{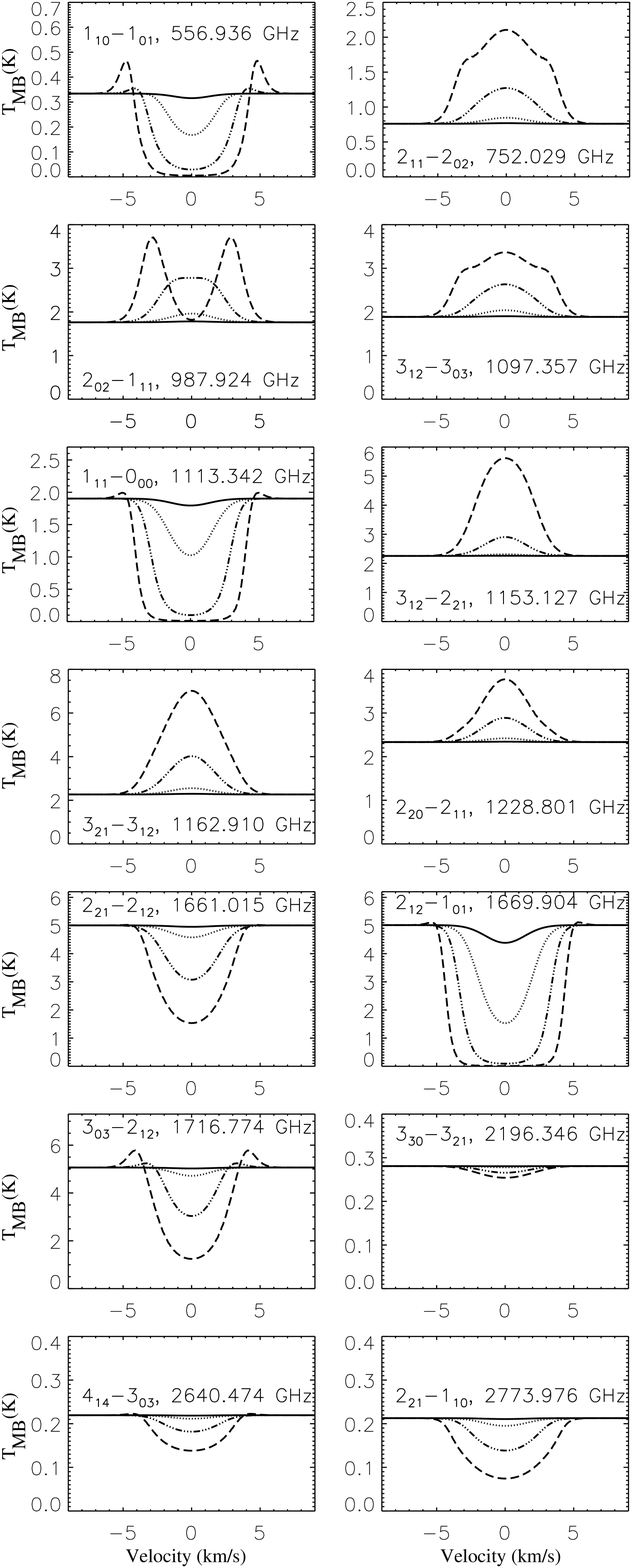}
\caption{Line profiles of ortho- and para-H$_\mathrm{2}$O transitions for models with $\alpha$\,=\,2. The water abundance, i.e., o-H$_\mathrm{2}$O\,+\,p-H$_\mathrm{2}$O, is uniformly distributed across the region and ranges from 10$^{-\mathrm{10}}$ to 10$^{-\mathrm{7}}$, with 10$^{-\mathrm{10}}$ ({\it solid}), 10$^{-\mathrm{9}}$ ({\it dotted}), 10$^{-\mathrm{8}}$ ({\it dash-dotted}), 10$^{-\mathrm{7}}$ ({\it dashed}).}
\label{fig:constant_alpha1_a20}
\end{figure}

We examine the influence of the adopted density and temperature distribution, i.e., adopting an $\alpha$\,=\,2 profile, on the resulting line profiles and strengths. In analogy to the models descibed in Section \ref{subsec:constant}, we assume the water abundance, ranging from $10^{-10}$ to $10^{-7}$, to be uniformly distributed. Resulting line profiles are plotted in Fig. \ref{fig:constant_alpha1_a20}, integrated brightness temperatures are listed in Table \ref{tab:constant_alpha1_a20}. 

In general, the line shapes are similar to the profiles presented in Section \ref{subsec:constant}. However, the 2$_\mathrm{21}$-2$_\mathrm{12}$, 3$_\mathrm{03}$-2$_\mathrm{12}$, and the ISO transitions are found in absorption, in contrast to the $\alpha$\,=\,1 models. This is a consequence of a steeper density slope, providing more material within the Herschel beam. Note that the optical depth in the lines increases by a factor of $\sim$\,5\,--\,10 for the high-lying transitions. The line strengths of the ground-state lines, i.e., 1$_\mathrm{10}$-1$_\mathrm{01}$ and 1$_\mathrm{11}$-0$_\mathrm{00}$, are similar to the strengths for the $\alpha$\,=\,1 models. This is not surprising since in the outer regions the temperature distribution follows the distribution of the $\alpha$\,=\,1 model. 

In addition, the effect of an $\alpha$\,=\,2 temperature and density profile on the line profiles and strengths is tested for models with an outflow cavity, described in Section \ref{subsec:cavity}. We find that the line shapes are similar to the profiles for the models with an $\alpha$\,=\,1 density gradient. However, the line strengths drop by a factor of $\sim$\,2 to 3 for the different transitions. 

\begin{table}[!t]
\label{tab:constant_alpha1_a20}
\centering
\begin{threeparttable}
\tabcolsep=1.6mm 
\caption{\small Integrated intensities, $\int${\it T}$_\mathrm{MB}$$\Delta${v} [K\,km\,s$^{-\mathrm{1}}$], for models with a constant H$_\mathrm{2}$O abundance. The adopted water abundance is denoted on top of each column. The density and temperature distribution is described by a $\alpha$\,=\,2 profile coefficient.}
\centering
\begin{tabular}{llrrrr}
\hline 
\hline 
Transition&Frequency&10$^{-\mathrm{10}}$&10$^{-\mathrm{9}}$&10$^{-\mathrm{8}}$&10$^{-\mathrm{7}}$\\
&[GHz]&&&&\\
\hline
\multicolumn{6}{c}{ortho-H$_\mathrm{2}$O transitions}\\
\hline
1$_\mathrm{10}$-1$_\mathrm{01}$ & 557.936  & $-$6.0($-$2)   &  $-$0.6      &  $-$1.6     & $-$2.1  \\
3$_\mathrm{12}$-3$_\mathrm{03}$ & 1097.35 &  5.6($-$2)   &   0.6      &   3.5     &  9.9 \\
3$_\mathrm{12}$-2$_\mathrm{21}$ & 1153.12 &  1.8($-$2)   &   0.2      &   2.4     &  15.4 \\
3$_\mathrm{21}$-3$_\mathrm{12}$ & 1162.91 &  0.1       &   1.0      &   7.0     &  23.7 \\
2$_\mathrm{21}$-2$_\mathrm{12}$ & 1661.01 &  $-$0.2      &   $-$1.6     &   $-$8.2    & $-$18.3  \\
2$_\mathrm{12}$-1$_\mathrm{01}$ & 1669.90 &  $-$2.3      &   $-$14.5    &   $-$31.9   &  $-$42.4 \\
3$_\mathrm{03}$-2$_\mathrm{12}$ & 1716.77 &  $-$0.1      &   $-$1.1     &   $-$6.4    &  $-$16.8 \\
3$_\mathrm{30}$-3$_\mathrm{21}$ & 2196.34 &  $-$1.1($-$3)  &   $-$1.0($-$2) &   $-$6.1($-$2)&  $-$0.1 \\
4$_\mathrm{14}$-3$_\mathrm{03}$ & 2640.47 &   $-$3.4     &   $-$3.1($-$2) &   $-$0.2    &  $-$0.4 \\
2$_\mathrm{21}$-1$_\mathrm{10}$ & 2773.97 &  $-$7.5($-$3)  &   $-$6.8($-$2) &   $-$0.4    &  $-$0.9 \\
\hline
\multicolumn{6}{c}{para-H$_\mathrm{2}$O transitions}\\
\hline 
2$_\mathrm{11}$-2$_\mathrm{02}$ & 752.029  &  3.1($-$2)   & 0.3   & 2.4   & 8.6\\
2$_\mathrm{02}$-1$_\mathrm{11}$ & 987.924  &  7.2($-$2)   & 0.7   & 5.2   & 8.7\\
1$_\mathrm{11}$-0$_\mathrm{00}$ & 1113.34 &  $-$0.4      & $-$3.2  & $-$10.1  & $-$14.5\\
2$_\mathrm{20}$-2$_\mathrm{11}$ & 1228.80 &  3.0($-$2)   &  0.3  & 2.2   & 6.3\\
\hline
\end{tabular}
N{\tiny OTE:} {\it a(b)} means {\it a}\,$\times$\,10$^\mathrm{\it b}$
\end{threeparttable}
\end{table}

\subsubsection{Opening angle}
In this section we briefly discuss the influence of the adopted outflow ope\-ning angle. We adopt density, temperature and abundance distributions from Section \ref{subsec:cavity}. Outflows with a cavity half-opening angle of 7.5$^{\circ}$ and 22.5$^{\circ}$ are computed. It is found that the adopted opening has minor influence on the line shapes. However, the line strengths drop (increase) by a factor of $\sim$\,2\,--\,3 (2) for models with an half-opening angle of 7.5$^{\circ}$ (22.5$^{\circ}$).
\subsubsection{Outflow water abundance}
In this section we briefly discuss the influence of an increased outflow water abundance by a factor of 10, i.e., 10$^{-\mathrm{6}}$. We adopt density, temperature and abundance distributions from Section \ref{subsec:cavity}. It is found that the adopted outflow water abundance has minor influence on the line shapes. However, the line strengths increase by a factor of $\sim$\,2 to 10. Note that the lowest gain in the line strengths is found for the ground-state transitions, i.e., 1$_\mathrm{10}$-1$_\mathrm{01}$, 2$_\mathrm{12}$-1$_\mathrm{01}$, and 1$_\mathrm{11}$-0$_\mathrm{00}$, and that the line strengths of the higher-lying transitions, e.g., 3$_\mathrm{12}$-3$_\mathrm{03}$, 3$_\mathrm{12}$-2$_\mathrm{21}$, 3$_\mathrm{21}$-3$_\mathrm{12}$, 2$_\mathrm{21}$-1$_\mathrm{01}$ and 2$_\mathrm{20}$-2$_\mathrm{11}$, are increased by a factor up to $\sim$10. This difference in enhancement factor can be explained due to the high gas and dust temperature in the outflow, favouring the excitation of the high-lying transitions.
 
\section{Summary and Conclusions}
We have constructed models to examine the excitation of water in the high-mass star-forming region AFGL\,2591. Depending on the adopted density, temperature and abundance structure, a completely different set of line profiles and strengths is found. Hence, the line profiles are very sensitive to the adopted physical and chemical structure. We have found that ({\it i}) the ground-state transitions 1$_\mathrm{10}$-1$_\mathrm{01}$, 2$_\mathrm{12}$-1$_\mathrm{01}$ and 1$_\mathrm{11}$-0$_\mathrm{00}$, with relatively low upper energy levels ($\lesssim$\,110\,K), become highly optically thick in the outer regions. The line profiles for these transitions, are mainly dominated by the emission from the outer regions, and are therefore not useful to put constraints on the water abundance in the inner regions. However, ({\it ii}) the emission from lines with higher upper energy levels is dominated by the emission originating in the inner regions, and are therefore useful to probe the water abundance in the warm inner regions. ({\it iii}) For models with an outflow cavity, the outflow feature (blue peak less strong than the red peak) is best seen in the ground-state transitions of o- and p-H$_\mathrm{2}$O. ({\it iv}) The influence of a moderate disk (few 100 AU in size) in the centre of the AFGL\,2591 region does not change the water line profiles and strengths within the Herschel beam. 

The Herschel mission will thus greatly help to understand the structure of high-mass protostellar objects, and consequently the formation process of high-mass stars. 

\begin{acknowledgements}
We thank Marco Spaans, Xander Tielens, Ewine van Dishoeck and Tim van Kempen for helpful discussions and suggestions which have improved the paper.
\end{acknowledgements}

\bibliographystyle{aa}
\bibliography{finalartikelIV}

\end{document}